\documentclass[aps,twocolumn,superscriptaddress,dvips]{revtex4}
\usepackage{feynmp}
\usepackage{amsmath}
\usepackage{amssymb}
\usepackage{epsfig}
\usepackage{graphicx}
\usepackage{subfigure}
\usepackage{hyperref}
\usepackage{bbm}
\usepackage{color}

\newcommand{\Rmnum}[1]{\expandafter\@slowromancap\romannumeral #1@}
\allowdisplaybreaks[3]

\begin{document}

\title{Reply to ``Comment on `Towards exact solutions of
superconducting $T_c$ induced by electron-phonon interaction' "}

\author{Guo-Zhu Liu}
\altaffiliation{Contact author: gzliu@ustc.edu.cn}
\affiliation{Department of Modern Physics, University of Science and
Technology of China, Hefei, Anhui 230026, P. R. China}
\author{Zhao-Kun Yang}
\affiliation{Department of Modern Physics, University of Science and
Technology of China, Hefei, Anhui 230026, P. R. China}
\author{Xiao-Yin Pan}
\affiliation{Department of Physics, Ningbo University, Ningbo,
Zhejiang 315211, P. R. China}
\author{Jing-Rong Wang}
\affiliation{Anhui Province Key Laboratory of Condensed Matter
Physics at Extreme Conditions, High Magnetic Field Laboratory of the
Chinese Academy of Sciences, Hefei, Anhui 230031, P. R. China}
\author{Xin Li}
\affiliation{Department of Modern Physics, University of Science and
Technology of China, Hefei, Anhui 230026, P. R. China}
\author{Hao-Fu Zhu}
\affiliation{CAS Key Laboratory for Research in Galaxies and
Cosmology, Department of Astronomy, University of Science and
Technology of China, Hefei, Anhui 230026, China}
\author{Jie Huang}
\affiliation{Department of Modern Physics, University of Science and
Technology of China, Hefei, Anhui 230026, P. R. China}

\begin{abstract}
In a series of papers, we have proposed a non-perturbative
field-theoretic approach to deal with strong electron-phonon and
strong Coulomb interactions. The key ingredient of such an approach
is to determine the full fermion-boson vertex corrections by solving
a number of self-consistent Ward-Takahashi identities. Palle (see
Phys. Rev. B 110, 026501 (2024), arXiv:2404.02918) argued that our
Ward-Takahashi identities failed to include some important
additional terms and thus are incorrect. We agree that our
Ward-Takahashi identities have ignored some potentially important
contributions and here give some remarks on the role played by the
additional terms.
\end{abstract}

\maketitle



Strong fermion-boson interactions cannot be treated by weak-coupling
perturbation theory. While the complete set of Dyson-Schwinger (DS)
equations are exact and in principle contain a major part of
interaction-induced effects, they are not closed and thus
intractable. In a series of papers \cite{Liu21, Pan21, Yang22,
Zhu22, Huang23}, we proposed a non-perturbative approach to
determine fermion-boson vertex corrections by solving a number of
self-consistent Ward-Takahashi identities (WTIs) and also applied
this approach to study strong electron-phonon and Coulomb
interactions in a few condensed matter systems.

In the preceding comment, Palle \cite{Palle} questioned the validity
of our approach and argued that the WTIs derived in our papers
\cite{Liu21, Pan21, Yang22, Zhu22, Huang23} are not satisfied by the
results obtained via perturbative calculations. Palle \cite{Palle}
further pointed out that the mistake arises from the ignorance of
the $z-z'$ dependence of the current vertex functions in our
manipulation of point-slitting technique. After considering the
$z-z'$ dependence, Palle \cite{Palle} shown that the modified WTIs
should contain some additional terms that cannot be expressed purely
in terms of the full fermion propagator $G(p)$, which signals the
invalidity of our approach.

We have actually analyzed the potential contribution of the $z-z'$
dependence, but eventually decided not to include it for the
following reason. The two points $z$ and $z'$ come from one point
$z$, and the limit $z \rightarrow z'$ must be taken at the end. If
an electron propagates between $z$ and $z'$, it must carry a
momentum $\mathbf{k}$ that is conjugate with
$\mathbf{z}-\mathbf{z}'$. As $z-z'$ approaches to zero, the absolute
value $|\mathbf{k}|$ should become extremely large so as to obey the
Heisenberg uncertainty principle. However, $|\mathbf{k}|$ should not
take large values because the low-energy properties of a quantum
many-fermion system are dominantly governed by the electrons excited
around the Fermi surface. It thus appeared to us that the inclusion
of an internal momentum $\mathbf{k}$ between $z$ and $z'$ would lead
to an inconsistency.

On the other hand, however, we agree with Palle that the current
vertex function $\Gamma_{t}(q,p)$ derived from our WTIs at one-loop
level are different from the one-loop result of $\Gamma_{t}(q,p)$
calculated by carrying out perturbative expansion. Therefore, our
previous analysis of the $z-z'$ dependence need to be re-examined
more carefully. The WTIs presented in Refs.~\cite{Liu21, Pan21,
Yang22, Zhu22, Huang23} are incorrect and should be properly
modified to accommodate the $z-z'$ dependence. Palle \cite{Palle}
argued that the WTIs should contain a number of additional terms
given by
\begin{eqnarray}
\Delta_{m}(q,p) = \int_{k}
\left(\xi_{\mathbf{p}+\mathbf{q}+\mathbf{k}} -
\xi_{\mathbf{p}+\mathbf{k}} - \xi_{\mathbf{p}+\mathbf{q}} +
\xi_{\mathbf{p}}\right)\tilde{\Gamma}_{m}(q,p,k),\nonumber
\end{eqnarray}
where $\tilde{\Gamma}_{m}(q,p,k)$ arises from a straightforward
Fourier transformation of
$\tilde{\Gamma}_{m}(z_{3}-z',z-z_{4},z-z')$. Although it is not
entirely clear to us how to define the internal interval of momentum
$\mathbf{k}$ in a self-consistent way, we agree that such additional
terms should be incorporated in the WTIs.

The contribution of additional terms might be small under suitable
conditions. Below are some remarks.

(1) For a one-dimensional interacting electron system, the electrons
have a linear dispersion $\xi_{\mathbf{p}}=v^{~}_{\mathrm{F}}
\left(|\mathbf{p}|-p^{~}_{\mathrm{F}}\right)$, where
$p^{~}_{\mathrm{F}}$ is Fermi momentum. In this case, the additional
terms $\Delta_{m}(q,p)$ vanish. Then the current vertex function can
be expressed purely in terms of $G(p)$, which reproduces the result
previously obtained by Dzyaloshinskii and Larkin \cite{Larkin}. For
two- and three-dimensional systems, the additional terms do not
necessarily vanish.

(2) The interfacial superconductivity of one-unit-cell
FeSe/SrTiO$_{3}$ system studied in Ref.~\cite{Liu21} is induced by
the interaction between electrons and optical phonons. Such an
interaction is strongly peaked at $\mathbf{q}=0$. In the limit
$\mathbf{q} \rightarrow 0$, the additional terms
\begin{eqnarray}
\Delta_{m}(q,p) = \int_{k} \frac{\mathbf{k}\cdot
\mathbf{q}}{m_{e}}\tilde{\Gamma}_{m}(q,p,k)
\end{eqnarray}
vanish, as long as $\tilde{\Gamma}_{m}(q,p,k)$ is analytic in
$\mathbf{q}$. Moreover, the Fermi-surface approximation, i.e.,
$\xi_{\mathbf{p}}= \xi_{\mathbf{p}^{~}_{\mathrm{F}}}=0$, was
employed in the numerical computations of $T_{c}$ in
Ref.~\cite{Liu21}. Under such an approximation, both
$\xi_{\mathbf{p}+\mathbf{q}}-\xi_{\mathbf{p}}$ and $\Delta_{m}(p,q)$
vanish. Thus, the numerical results of the transition temperature
$T_{c}$ reported in Ref.~\cite{Liu21} are not expected to be changed
by the additional terms.

(3) For fermion-boson interacting systems that are not dominated by
zero-$\mathbf{q}$ forward scattering, additional terms are small
only under the approximation
$\xi_{\mathbf{p}}=\xi_{\mathbf{p}^{~}_{\mathrm{F}}}=0$. If such an
approximation is not justified, the contribution of additional terms
cannot be ignored.

(4) When electrons are coupled to two sorts of bosons, the DS
equations become much more complicated than the case of a single
fermion-boson coupling owing to the presence of four different
interaction vertex functions, as shown by Eq.~(27) in
Ref.~\cite{Yang22} and Eq.~(40) in Ref.~\cite{Huang23}. The
identities satisfied by the interaction and current vertex functions
given by Eqs.(40-41) in Ref.~\cite{Yang22} and Eq.(50) and Eq.(52)
in Ref.~\cite{Huang23} are still valid and can be used to simplify
the DS equation of $G(p)$. One can see from Eq.(42) in
Ref.~\cite{Yang22} and Eq.(53) in Ref.~\cite{Huang23} that the
simplified DS equation of $G(p)$ contains, apart from $G(p)$ and
three free propagators, merely one single current vertex function.
Even though such a current vertex function cannot be determined
rigorously, one could compute it approximately by means of series
expansion if the system has a small coupling parameter.

In summary, we admit that the DS equations of the full fermion
propagator $G(p)$ derived in Refs.~\cite{Liu21, Pan21, Yang22,
Zhu22, Huang23} are not self-closed once the $z-z'$ dependence of
current vertex functions $\tilde{\Gamma}_{m}(z_{3}-z',z-z_{4},z-z')$
is taken into account. The additional terms appearing in modified
WTIs are small under certain approximations, but generically cannot
be neglected. The structure of the additional terms remains unknown
and need to be carefully investigated. The modified WTIs impose
exact constraints on current vertex functions. Such constraints do
not suffice to make the DS equation of $G(p)$ self-closed, but
provide useful guidance for the exploration of suitable approximate
form of current vertex functions.

We thank G. Palle for insightful comments.

\end{document}